\documentclass[a4paper,11pt]{article}
\usepackage{jinstpub} 
\usepackage[latin9]{inputenc}
\usepackage{float}
\usepackage{subfig}
\usepackage{esint}
\usepackage{xcolor}
\usepackage{svg}
\usepackage{siunitx}
\usepackage{orcidlink}

\title{MEOP based \textsuperscript{3}He polarization and injection system for experiments below 1K}

\author[a]{T. Rao}
\author[b]{L. Barr\'on-Palos}
\author[a]{I. Berkutov}
\author[c]{C. Crawford}
\author[a]{R. Golub}
\author[a]{P. Huffman}
\author[d]{M. Konieczny}
\author[a]{E. Korobkina}
\author[e]{Austin Reid\,\orcidlink{0000-0003-1741-2223}}
\author[b]{B. Salazar-\'Angeles}
\author[c]{C. Smith}
\author[a]{R. Tat}
\author[b]{T. Zanatta-Mart\'inez}
\affiliation[a]{Department of Physics, North Carolina State University, Raleigh, NC 27695, U.S.A.}
\affiliation[b]{Instituto de F{\'i}sica, Universidad Nacional Aut{\'o}noma de M{\'e}xico, Apartado Postal 20-364, 01000, Mexico}
\affiliation[c]{Department of Physics and Astronomy, University of Kentucky, Lexington, KY 40506, U.S.A.
}
\affiliation[d]{Department of Physics and Astronomy, Ithaca College, Ithaca, NY 14850, U.S.A}
\affiliation[e]{Department of Physics, Tennessee Tech University, Cookeville, TN 38505, U.S.A}
\emailAdd{trao@ncsu.edu}

\abstract{Metastability exchange optical pumping (MEOP) is a widely used technique for producing polarized \textsuperscript{3}He. In connection with an experiment to search for the electric dipole moment of the neutron (nEDM)  we have built  a MEOP based \textsuperscript{3}He polarization and injection system to prepare 80 \% polarized \textsuperscript{3}He at room temperature which will be injected into a ~400 mK measurement cell filled with superfluid \textsuperscript{4}He. We describe the polarization and injection system, which is designed to allow for final concentrations of $10^{-8}-10^{-10}$ of 80 \% polarized \textsuperscript{3}He in the superfluid filled measurement cell. Only $\approx0.72\%$ polarization loss due to gradients is expected during injection}

\begin{document}
\maketitle
\flushbottom

\section{Introduction}
One of the most common ways of polarizing \textsuperscript{3}He is metastability exchange optical pumping (MEOP). In MEOP, an RF discharge is used to excite \textsuperscript{3}He atoms into a metastable state which can be polarized via optical pumping by a laser. This polarization is then transferred to ground state atoms through metastability-exchange collisions. MEOP is typically performed at pressures around \SI{1}{\milli\bar}. See \cite{Gentile} for a general review of \textsuperscript{3}He optical pumping methods such as MEOP.

The polarized \textsuperscript{3}He produced by MEOP has applications in a variety of fields from medical magnetic resonance imaging to axion searches ~\cite{Gentile}. 
In a high precision  neutron electric dipole moment (nEDM) searches polarized \textsuperscript{3}He found its use in magnetometers \cite{Kraft,borisov2000feasibility}.
In nEDM experiments co-magnetometry is used to understand the field gradients in the measurement cell and correct a Larmor precession signal for the magnetic field instabilities~\cite{Abel}.  Mentioned above magnetometers operate at room temperature.  A low temperature polarized \textsuperscript{3}He magnetometer has been used to measure the field gradients inside cryostats \cite{Fan}.

MEOP has additionally been used in NMR studies of the spin dynamics of liquid \textsuperscript{3}He--\textsuperscript{4}He mixtures as well as pure \textsuperscript{3}He. The work of \cite{Hayden1} involves a room temperature \textsuperscript{3}He polarization cell connected to a loop that goes down to a temperature below \SI{1}{\kelvin} which is partially filled with a \textsuperscript{3}He--\textsuperscript{4}He mixture. A heater is used to drive the circulation \textsuperscript{3}He keeping a constant concentration of \textsuperscript{3}He in the \textsuperscript{4}He (a few percent) while the cell maintains \textsuperscript{3}He polarization. A similar method was used in \cite{Hayden2} where a room temperature cell is connected to a \SI{1}{\kelvin} NMR cell and the concentration of \textsuperscript{3}He is controlled using a mass flow controller. This system is capable of maintaining polarized \textsuperscript{3}He concentrations of a 1-5 \%. 

Injection methods of delivering room temperature polarized \textsuperscript{3}He produced via MEOP into a sub \SI{1}{\kelvin} measurement cell for NMR studies have also been achieved. In \cite{Tastevin1}\cite{Tastevin2} a closed pure \textsuperscript{3}He (or mixed \textsuperscript{3}He--\textsuperscript{4}He) system connecting room temperature polarization cell and a measurement cell submerged in a liquid \textsuperscript{3}He bath is used. Cooling the \textsuperscript{3}He bath from \SI{1.5}{\kelvin} to \SI{400}{\milli\kelvin} condenses the \textsuperscript{3}He into the measurement cell. The \textsuperscript{3}He polarization delivered to the measurement cell was \SI{50}{\percent} or less. While \cite{Yoder} describes a system where \textsuperscript{3}He is polarized at room temperature, the injected into a cell filled with liquid \textsuperscript{3}He. This system can achieved a minimum concentration of approximately $10^{-4}$, with polarization estimated to be 10\%. The minimum concentration is set by pressure of the polarization cell needed for MEOP, volume of the superfluid filled measurement cell. The entire experiment is enclosed in a uniform holding field. 

Additionally, low temperature NMR of polarized \textsuperscript{3}He is used in the nEDM experiment design described in \cite{golub94}. This experiment utilizes a cryogenic \textsuperscript{3}He co-magnetometer diluted in superfluid liquid \textsuperscript{4}He at temperatures below \SI{500}{\milli\kelvin}. In this design the co-magnetometer doubles as a detection mechanism for the neutron spin precession.  An apparatus to measure the nEDM using a cryogenic  environment of polarized \textsuperscript{3}He and neutrons in superfluid \textsuperscript{4}He is described in \cite{Ahmed}. The described experiment aims at an unprecedented ultimate statistical uncertainty of \num{3e-28}. As part of this proposal a Systematic and Operations Studies (SOS) apparatus is being developed at the Triangle Universities Nuclear Laboratory (TUNL). This apparatus is a scaled down version of the main experiment apparatus without the high voltage systems needed to measure an EDM. The main design goal of SOS apparatus is to allow precise and effective spin manipulation of both polarized \textsuperscript{3}He and neutrons in a highly homogeneous weak magnetic field of \SI{3}{\micro\tesla} using NMR. Such apparatus  will allow us to verify the experimental techniques required for the proposed  nEDM measurement and to study different phase shifting systematic effects. MEOP technique will be used to produce the polarized \textsuperscript{3}He for the SOS.

\subsection{Overview of the MEOP assembly design}

The MEOP based \textsuperscript{3}He polarization system of SOS apparatus is designed to be capable of  producing \SI{80}{\percent} polarized \textsuperscript{3}He at room temperature and delivering the polarized gas into a superfluid filled measurement cell of the SOS apparatus at temperature range of \SI{0.3}{\kelvin} -- \SI{0.5}{\kelvin} and a wide range of relative concentrations from \num{e-8} to \num{e-10}.  The entire MEOP assembly is enclosed inside a cosine theta holding field magnetic coil. The present publication describes in detail design and testing of the gas handling system and \textsuperscript{3}He friendliness of the pneumatic valves (Sections 2 and 3); as well as the design and results of field mapping of the magnetic coil (Section 4). Injection of  polarized \textsuperscript{3}He into the superfluid helium and procedures for preserving polarization during injection are described in Section 5. This system is unique in that we can achieve ultra low concentrations where the dynamics of the \textsuperscript{3}He is completely governed by interactions with the phonons. At the higher concentrations studied in \cite{Yoder} and others \textsuperscript{3}He--\textsuperscript{3}He interactions are relevant. Additionally, the work in \cite{Yoder} was limited to low polarization, while the work described in \cite{Tastevin1, Tastevin2} was limited to polarizations up to \SI{50}{\percent}. Due to the low concentrations being studied at the SOS the NMR signal will be comparatively quite small and it will be critical to have has high as concentration as possible to maximize our signal to noise. 
The design of this apparatus presents us with the unique challenge of a developing high polarization (\SI{80}{\percent}) room temperature polarization system capable of delivering ultra low concentrations of \textsuperscript{3}He ($10^{-10}$) into a superfluid measurement cell with minimal loss of polarization during injection. To achieve these goals three "dilution volumes" can be used to reduce the amount of \textsuperscript{3}He injected into the measurement cell. This allows measurements at the $10^{-8}$, $10^{-9}$, and $10^{-10}$ concentration level in the \SI{3}{\liter} measurement cell. $3 \times 10^{-10}$ will be \textsuperscript{3}He concentration in the proposed nEDM experiment. Additionally, the system is designed so that after polarization the \textsuperscript{3}He can be pressured with \textsuperscript{4}He preventing depolarization between dilution stages and injection. Finally, the gas handling system and holding field used for polarizing the gas is designed to allow for $T_{1}>1000$ s in the MEOP cell.
\begin{figure}[H]
    \centering
    \includegraphics[scale=0.45]{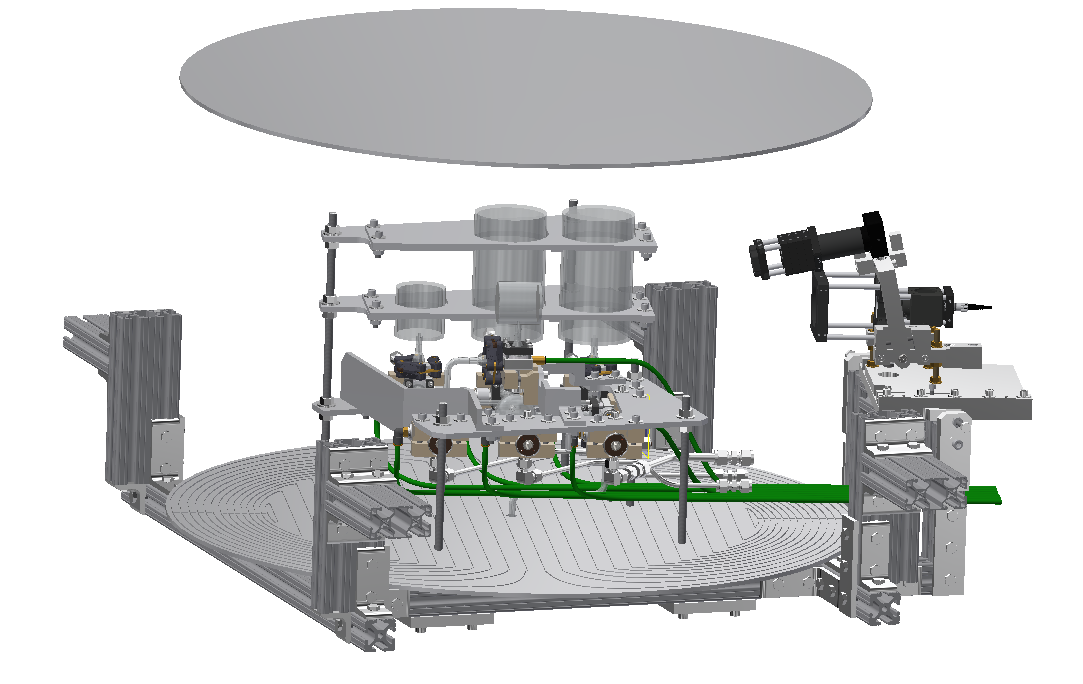}\caption{Model of MEOP coil, manifold, dilution volumes, and laser optics. The main body of the coil is rendered invisible to make manifold visible. Only the top and bottom of the coil are visible.}
    \label{fig:MEOPManifoldCoil}
\end{figure}
Figure \ref{fig:MEOPManifoldCoil} shows a model of the room temperature polarization and injection system as well as the optical pumping laser optics, polarimeter, and top and bottom end caps of the holding field coil that encloses it. The main body of the coil is not shown so that the manifold is visible. The laser is a 1083 nm CYFL-GIGA series Ytterbium-doped fiber laser. The polarimeter operates by measuring the circular polarization of the 668 nm fluorescence light from the polarization cell.

\section{\texorpdfstring{\textsuperscript{3}}{3}He gas handling system design}
\begin{figure}[H]
    \centering
    \includegraphics[scale=0.5,trim={0 5cm 0 0},clip]{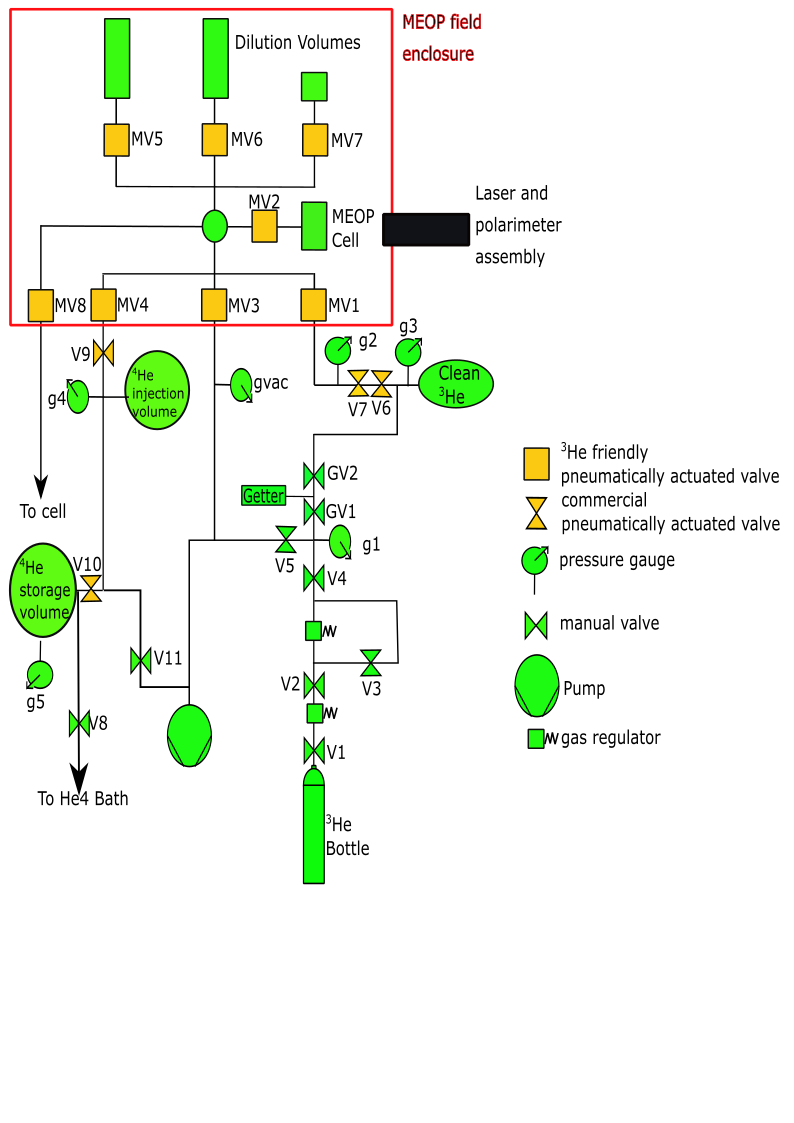}\caption{Schematic Diagram of \textsuperscript{3}He GHS. Gas from the \textsuperscript{3}He bottle is reduced to approximately 100 mbar using a pair of regulators and gettered for several hours. The gettered gas is then used to fill the clean \textsuperscript{3}He volume to 8 mbar. v7, v6, MV1, and MV2 are used to fill the MEOP cell to 1.3 mbar. The 1.25 cc volume between v7 and v6 allows for the MEOP cell to be filled in increments as small as of 0.06 mbar when the clean volume is full. The \textsuperscript{4}He injection volume is used to pressurize the MEOP cell before dilution to 500 mbar, using gas from the \textsuperscript{4}He storage volume. The three dilution volumes are used to reduce the amount of \textsuperscript{3}He injected into the measurement cell.}
    \label{fig:GHSfigure}
\end{figure}

The \textsuperscript{3}He gas handling system (GHS) is shown in Figure \ref{fig:GHSfigure}. It is used to purify, polarize, and reduce the amount of \textsuperscript{3}He to be injected, in order to achieve the desired concentration of polarized \textsuperscript{3}He in the superfluid filled measurement cell. In addition the GHS must provide a means of pressurizing the \textsuperscript{3}He with \textsuperscript{4}He before dilution so the pressure dependent $T_{1}$ is long enough that the \textsuperscript{3}He doesn't depolarize during the dilution process or injection. The higher density of \textsuperscript{4}He also helps ensure the injected \textsuperscript{3}He condenses in the superfluid. The GHS also must not produce any long or short range gradients that could depolarize the polarized \textsuperscript{3}He.

\begin{figure}[H]
    \centering
    \subfloat[]{\includegraphics[scale=0.55]{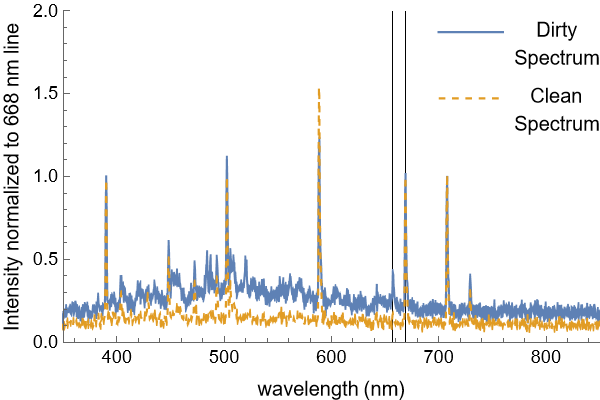}}\hspace{5mm}
    \subfloat[]{\includegraphics[scale=0.55]{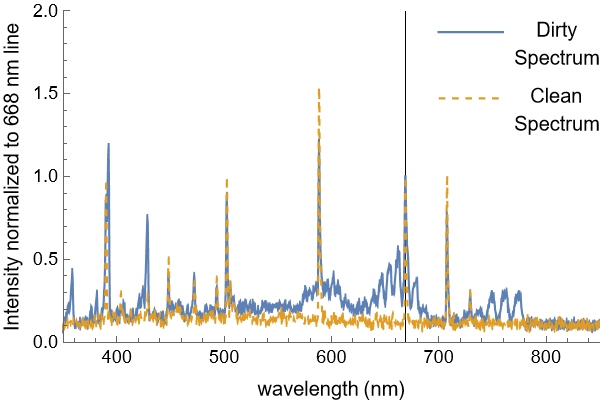}}

    \caption{Examples of clean vs dirty \protect\textsuperscript{3}He spectrum. (a) the small marked peak at 656 nm in the dirty spectrum corresponds to the n=3 to n=2 transition in the Hydrogen. Most likely from water vapor trapped in the system cracking on the getter. (b) Multiple non-Helium peaks over a wide range of wavelengths can be seen in the dirty spectrum. Indicative of a small leak to air. The clean spectrum shown on on both plots was taken after gettering. The intensities plotted on the y axis are relative to the intensity of the 668 nm line. }
    \label{fig:Spectrum}
\end{figure}

The gas handling system is designed to purify approximately 16 injections worth of \textsuperscript{3}He using a non evaporation getter (SAES model S5H0380 MAP/7/20 mini CFF). The \textsuperscript{3}He gas is initially pressurized in a 20 standard liter bottle. Two regulators are used to reduce the pressure to approximately 100 mbar at gauge g1. With v4 and GV2 closed, the space between valves v4 and GV2 is gettered. Depending on how clean the gas and GHS initially is, it can take hours to properly getter the gas. Analyzing the spectrum of the gettered gas, by applying an RF discharge to the polarization cell, when filled with \textsuperscript{3}He, can be used to determine the cleanliness of the gas. Examples, of the spectrum of clean vs dirty gas can be seen in figure\ref{fig:Spectrum}. These spectra were measured by applying a RF discharge to 1 Torr of gas in a 80 cc volume. An OceanView spectrometer was then used to measure the resulting spectrum. The peak at 668 nm is of particular interest since it is used to measure the polarization of the gas. A peak at 656 nm, corresponds to the n=3 to n=2 transition in hydrogen, most likely coming from water vapor in the system.

After gettering, the clean gas is let into the \SI{0.5}{\liter} clean \textsuperscript{3}He storage volume. Clean gas can be let into the polarization cell from the clean volume by utilizing valves v7, v6, MV1, and MV2. Gauge g2 can be used to monitor the pressure at the cell, and the small space between v6 and v7 (1.25 cc) allow gas to be added with a precision of 0.06 mbar when the clean volume is full. Optimally we intend to polarize with approximately 1.3 mBar in the MEOP cell. 

Once the polarization cell is filled with \textsuperscript{3}He, it can be polarized using MEOP. In MEOP polarization is accomplished using a 10 MHz RF discharge to excite ground state electrons to the 2S metastable state. From there optically pumping the C8 transition is used to polarize the metastable state. Exciting this transition requires 1083 nm light laser light. A 10 W Ytterbium doped fiber laser was used to excite the C8 transition. A Laser expander (seen along with the polarimeter in Figure 4) is used to maximize the laser light's coverage of the cell. Collisions between the polarized metastable atoms and unpolarized ground state atoms result in nuclearly polarized ground state atoms. See \cite{Batz} for a detailed description. The polarization is determined by measuring the polarization of 668 nm florescence light from the cell using a polarimeter based on the design described in \cite{Maxwell}. 

The linear polarizer in the polarimeter can be adjusted to block any linearly polarized light reflected off the cell. However, a significant amount of unpolarized laser light is still detected resulting in a measurable background. Increasing the RF amplitude will improve the signal to noise, however, higher RF has been observed to reduce polarization. Early polarization measurements focused demonstrating high polarization (such as those taken for figure 7) where taken with a choice of RF to maximize polarization. Later measurements (like the $T_{1}$ measurements summarized in tables 1 and 2) were generally taken with larger RF voltages to maximize our signal to noise at the cost of reducing maximum polarization to (73-75 \%). We perform MEOP in a 5 G holding field. The coil generating the field encompasses the MEOP cell, glass manifold, and dilution volumes  (see figure \ref{fig:MEOPManifoldCoil}).
 In order to achieve a $T_{1}>1000$ s the gradient must be less than 0.0833\%/cm at the cell. See Section 4 for a detailed description of the coil.

Injecting all the polarized gas in the cell would result in a concentration of $4.7\times10^{-8}$ in the \SI{3}{\liter} superfluid filled measurement cell. \textsuperscript{3}He must be removed from the system before injecting in order to achieve a lower final concentration. However, the gas is first pressurized with  \textsuperscript{4}He to 500 mbar from a \SI{0.3}{\liter} \textsuperscript{4}He injection volume, which in turn is filled from the \SI{3.8}{\liter} \textsuperscript{4}He storage volume, which contains approximately 1000 mbar of clean helium from a LHe dewar. The purpose of pressurizing the \textsuperscript{3}He in the MEOP cell is to increase the $T_{1}$ of the gas, and to help ensure the polarized \textsuperscript{3}He condenses in the superfluid filled measurement cell during injection. From \cite{McGregor,Cates} we see the $T_{1}$ of the gas due to depolarizing gradients is given by

\begin{equation}
    \frac{1}{T_{1,\mathrm{grad}}}=\frac{(\nabla B_{r})^{2}}{B_{0}^{2}}\frac{D}{1+\gamma^{2}B_{0}^{2}\frac{D^{2}M^{2}}{k_{B}^{2}T^{2}}}
    \label{1}
\end{equation}

where $B_{r}$ is the field transverse to the holding field $(B_{0}$), $\gamma$ is the gyromagnetic ratio, $M$ is the mass of a \textsuperscript{3}He atom, and the diffusion constant given by $D=D_{0}P_{0}/P_{cell}$. $P_{cell}$ is the pressure in the polarization cell, $P_{0}=1$ Torr is a reference pressure, and $D_{0}$ is the diffusion constant at the reference pressure. From \cite{Barbe} $D_{0}=1440\pm80$ cm\textsuperscript{2}/s for \textsuperscript{3}He self diffusion, or $1320\pm50$ in the case of  \textsuperscript{3}He--\textsuperscript{4}He diffusion. In the case where $\gamma^{2}B_{0}^{2}\frac{D^{2}M^{2}}{k_{B}^{2}T^{2}}\ll1$ or equivalently $P\gg0.0135$ Torr for a 5 G holding field (0.027 Torr for 10 G) eq. (\ref{1}) can be rewritten as

\begin{equation}
    T_{1,\mathrm{grad}}=\frac{B_{0}^{2}}{(\nabla B_{r})^{2}}\frac{1}{D}
    \label{2}
\end{equation}

Pressures of about $10^{-2}$ Torr will result in short $T_{1}$ times of about 5 s that will make maintaining polarization during injection impossible. Pressurizing with 500 mbar of \textsuperscript{4}He increases the $T_{1}$, preventing depolarizing during dilution and injection. To reduce the concentration MV2 and MV5 are quickly opened and then closed, filling a dilution volume, and reducing the pressure in the polarization cell to $P_\mathrm{cell}=51$ mbar ($T_{1,\mathrm{grad}}=39000$ s). The gas in the manifold is then pumped out and the cell is repressurized to 500 mbar. This leaves $3\times10^{17}$ atoms of \textsuperscript{3}He in the cell, which if injected will result in a final concentration of $4\times10^{-9}$. Repeating this procedure using the remaining two dilution volumes results in final concentrations of $3\times10^{-10}$ and $1\times10^{-10}$ respectively. Note the third dilution volume $V_{\mathrm{Dil},3}$ is 110 cc nominal (105 cc actual), while the other two are 670 cc nominal (655 cc actual). The manifold ($V_\mathrm{man})$is 56 cc. After dilution the gas can be injected by opening MV2 and MV3. The resulting concentration in the \SI{3}{\liter} measurement cell after the ith round of dilution can be calculated using the following recursion relation to find the number of \textsuperscript{3}He atoms in the cell 

\begin{equation}
    N_{3,i}=N_{3,i-1}\left(\frac{V_\mathrm{cell}}{V_\mathrm{cell}+V_\mathrm{man}+V_{\mathrm{Dil},i}}\right)^{i}
    \label{3}
\end{equation}

along with $x_{3,i}=N_{3,i}/(N_{3,i}+N_{4})$ to calculate the concentration. Where $N_{4}$ is the number of \textsuperscript{4}He atoms in the \SI{3}{\liter} measurement cell, which can be calculated from the density of He-II at 0.4 K. The initial number of \textsuperscript{4}He atoms in the polarization cell before injection is $N_{3,0}=\frac{P_\mathrm{cell}V_\mathrm{cell}}{kT}$.

\section{Testing \texorpdfstring{\textsuperscript{3}}{3}He friendliness of the pneumatically actuated valves}

\begin{figure}[H]
    \centering
    \includegraphics[]{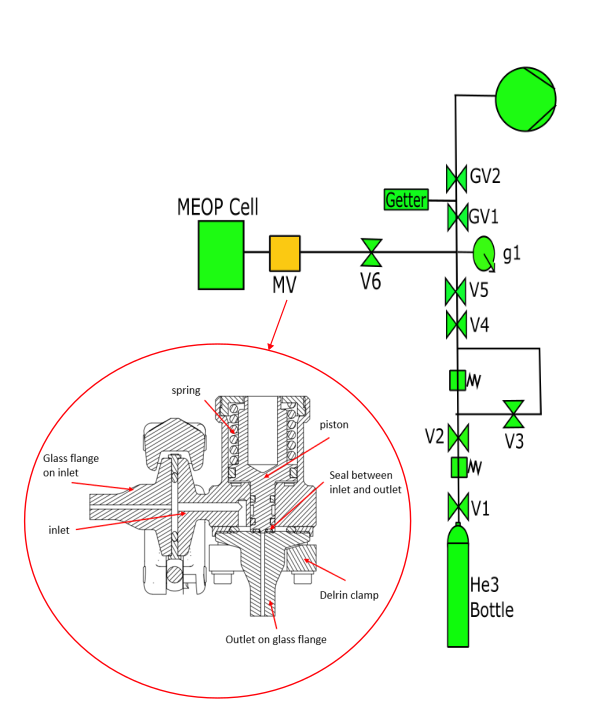}
    \caption{V1 though V5 are used to prepare $\approx 1$ Torr of gas in the MEOP cell. The cell is left open to gettering by having GV1 open for 2 hrs to clean the gas. MV is the pneumatic valve under test, which is located $\approx 3$~in away from the cell. The outlet of the valve is connected to the cell, while the inlet connects to the rest of the GHS}
    \label{fig:TestSystem}
\end{figure}

As mentioned in the previous section to achieve a $T_{1}>1000$ s a magnetic field gradient less than \SI{0.0833}{\percent/\centi\meter} is required. This requirement is met by the holding field produced by a double cosine theta coil that generates the holding field for MEOP (see section 4). However, additional gradients can be introduced into the system from magnetic impurities in the GHS system itself. Of particular concern are the eight pneumatically actuated valves (MV1-MV8) inside the magnetic field region that control the flow of polarized \textsuperscript{3}He and the \textsuperscript{4}He used to pressurize the cell. These pneumatically actuated valves are made of Al5083 which has a maximum iron impurity of \SI{0.4}{\percent} \cite{Azom}. The designs of these valves are based on the valves used for a MEOP based \textsuperscript{3}He system polarization system for neutron polarization filters at the Institut Laue-Langevin (ILL) \cite{Heil,Jullien}. Figure \ref{fig:TestSystem} includes a schematic drawings of a valve. The inlet and outlet of the valve are separated by the o-ring seal on the piston. 70~psi of compressed air is required to compress the spring and open the valve. The valve closes when the pressure is released. The Delrin clamp applies pressure to make the seal at the piston between inlet and outlet as well as the seal with the large diameter o-ring between the valve and air. The version of the valve pictured uses a KF style connection between the inlet and a glass flange. An alternate style valve is also used where the inlet is a straight tube used to mage a swagelok connection to the Al tubing on the GHS.

To ensure the valves used on the glass manifold inside the MEOP coil don't introduce magnetic field gradients, or gas contaminants, and are therefore \textsuperscript{3}He "friendly", tests were conducted with a simpler MEOP gas handling system. In the this setup $\approx 1$~Torr of gas is prepared, gettered and then polarized, without a system for diluting or storing gas. Figure \ref{fig:TestSystem} shows the layout of the test apparatus

The laser, optics, and polarimeter are the same as in the actual SOS apparatus. Although for this test the laser was only operated at \SI{7}{\watt} instead of the laser's maximum output of \SI{10}{\watt}. Like in the SOS apparatus beam expander optics are used to maximize the cross sectional area of the cell covered by laser light.  The magnetic field in the test apparatus however is significantly less uniform. Consequently, to get a $T_{1} \approx 1000$ s a holding field of 10 G is required. Initial tests were conducted with a glass stopcock to provide a baseline measurement for the system. Later measurements replaced this glass stopcock with the \textsuperscript{3}He friendly pneumatically actuated valve under test, allowing us to determine if the Al valves introduce any gradients compared to an all glass system.

\subsection{Initial glass stopcock test}

\begin{figure}[H]
    \centering
    \includegraphics[scale=0.6]{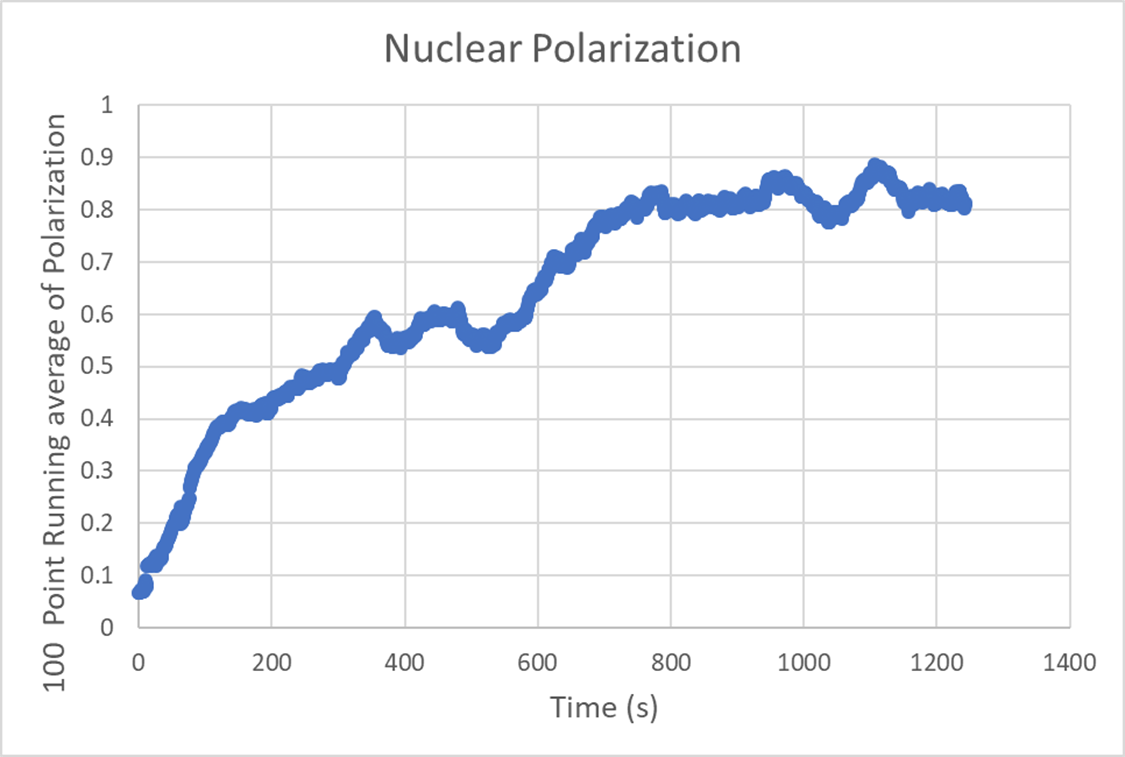}
    \caption{100 point running average of the measured polarization as a function of time. Raw polarization data is measured at a rate of 2 Hz. }
    \label{fig:80Polarization}
\end{figure}

With the glass stopcock installed, 0.86 Torr of gas was injected into the MEOP cell. The MEOP cell was then closed off the rest of the test system by closing the stopcock and the gas was polarized. Figure \ref{fig:80Polarization} shows a typical polarization curve

\begin{figure}[H]
    \centering
    \includegraphics[scale=0.6]{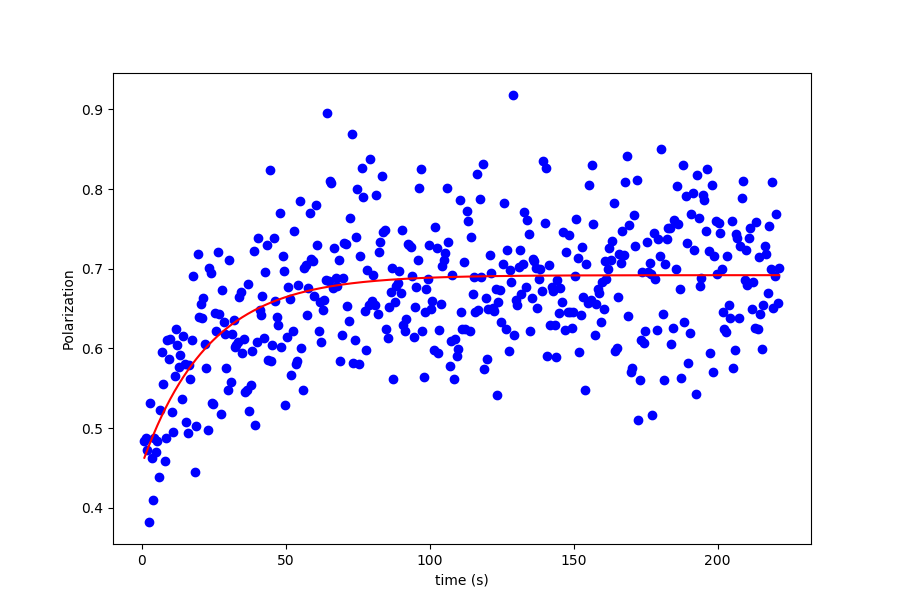}
    \caption{Example curve of gas repolarizing after the RF was turned off for 296 s. RF is turned on at t=0 s. The red curve is a fit to the data using $A(1-e^{-t/\tau})+y_{0}$ }
    \label{fig:polarizing}
\end{figure}

After achieving 80\% polarization, a series of $T_{1}$ measurements were made. The $T_{1}$ measurements were taken with the RF off otherwise depolarization from the plasma dominates the the signal. With the RF off the signal $T_{1}$ is dominated by long range magnetic field gradients. The contribution from wall depolarization can also be measured by taking measurements at multiple field strengths and extracting the component of the $T_{1}$ that is independent of the magnetic field. $T_{1}$ measurements are taken by polarizing the gas, then turning of the RF. During this time the gas depolarizes. After waiting for a set amount of time the RF is turned back on and the new polarization is measured. The gas is then allowed to repolarize and the process is repeated for multiple depolarization times. In order to extract the polarization of the gas immediately after turning the RF back on the the re-polarizing signal is fit to $A(1-e^{-t/\tau})+y_{0}$; where $t$ is the time after turning the RF back on, $y_0$ is the polarization of the gas at $t=0$, $A$ is the change in polarization while polarizing the gas for the next measurement, and $\tau$ is the characteristic  timescale for polarizing the gas. Figure \ref{fig:polarizing} shows an example of fitting the repolarizing gas.

In order to compare the different measurements each re-polarization curve is normalized  by the polarization before the RF was turned off. The normalized polarization for the i\textsuperscript{th} measurement is then

\begin{equation}
\frac{y_{0,i}}{A_{i-1}+y_{0,i-1}}.
\end{equation}

\begin{figure}[H]
    \centering
    \includegraphics[scale=1]{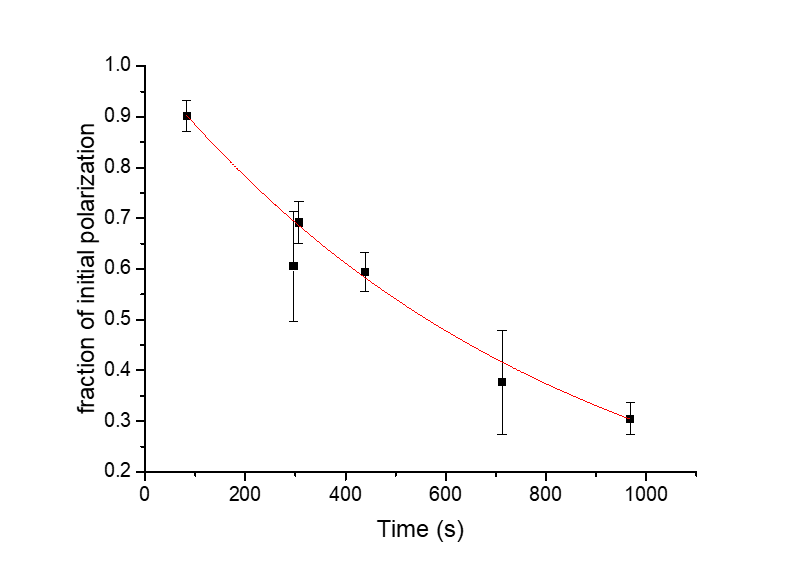}\caption{Normalized polarization as a function of depolarization time for 0.86 Torr of gas with a holding field of 10 G. $T_{1}$ is measured by fitting the data to a decaying exponential. $T_{1}=810\pm22$ s}
    \label{fig:T1}
\end{figure}

Below is a plot of the polarization as a function of depolarization time. Figure \ref{fig:T1} shows the normalized polarization as a function of the depolarization time ($t_{d}$). The $T_{1}$ is determined by fitting to a decaying exponential, $e^{-t_{d}/T_{1}}$ .

Based on this measurement if the measured $T_{1}$ is assumed to be completely due to magnetic gradients a value of $0.0086\pm0.0003$ G/cm is obtained for the gradient by using eq.~(\ref{2}). At \SI{0.86}{Torr} the diffusion constant is approximately \SI{1670}{\centi\meter^2/s}. If we combine eq.~(\ref{2}) with the effect of wall losses the effective $T_{1}$ is now given by 

\begin{equation}
    T_{1}=\left(\frac{1}{T_{1,\mathrm{wall}}}+\frac{1}{T_{1,\mathrm{grad}}}\right)^{-1}=\left(\frac{1}{T_{1,\mathrm{wall}}}+\frac{(\nabla B_{r})^{2}}{B_{0}^{2}}D\right)^{-1}
    \label{4}
\end{equation}
    
\begin{figure}[H]
    \centering
    \includegraphics[scale=1]{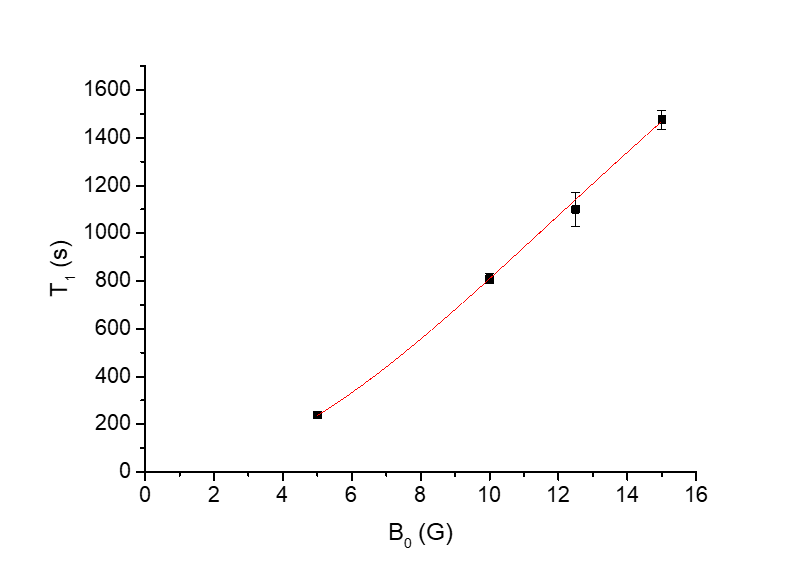}\caption{$T_{1}$ as a function of magnetic field for 0.86 Torr of gas. Fitting gives $T_{1,\mathrm{wall}}=4170\pm260$ s and $\nabla B_{r}=0.00772\pm0.00025$ G/cm}
    \label{fig:T1wall}
\end{figure}

The wall $T_{1}$ ($T_{1,\mathrm{wall}}$) can therefore be determined by measuring the $T_{1}$ for different magnetic fields and fitting the result to $T_{1} (a+b/B_{0}^{2})^{-1}$; where $a=1/T_{1,\mathrm{wall}}$ and $b=(\nabla B_{r})^{2}D$. Figure \ref{fig:T1wall} shows the results of this fitting for the glass stopcock data. The portion of  $\nabla B_{r}$  proportional to $B_0$  will be included in  $T_{1,\mathrm{wall}}$ 

Doing so gives $T_{1,\mathrm{wall}}=4170\pm260$ s and $\nabla B_{r}=0.00772\pm0.00025$ G/cm. Note that this measurement not only provides a measurement of the wall depolarization but also a correction to the the previously quoted gradient that ignored the contribution of the walls to the $T_{1}$. Subsequent measurements with the Al valves were taken at 1.0 Torr. If the $T_{1}$ is assumed to be purely from gradients scaling the 0.86 Torr $T_{1}$ to 1.0~Torr gives $940\pm30$~s. If the $T_{1}$ at 1.0~Torr is instead calculated from eq. (\ref{4}) we instead have $910\pm30$~s.

\subsection{Tests of MEOP valves with KF style flanges}

After these tests the glass stopcock was removed from the system and glass flanges were added to allow for installing the pneumatically actuated valves. The magnetic field region of the GHS includes 5 valves as shown in Table 1. One of these valves came from ILL. $T_{1}$ measurements we conducted in the same way as for the glass stopcock resulting in the following values

\begin{table}[H]
    \caption{Measured $T_{1}$ for the 5 MEOP valves with KF style flanges. Measurements were taken at 10 G for all 5 valves. For valves 2-4 additional measurements were made at 5, 12.5, and 15 G, except valve 4 which is missing a 15 G measurement, which were used to extract values for the gradient and wall $T_{1}$. The $T_{1}$ at 10 G of 4 out of 5 valves are consistent with the $T_{1}$ with the Glass stopcock installed. Valve 1 which has a 22\% lower $T_{1}$ is still acceptable for our purposes. The pressure in the cell for all tests was 1.0 Torr}
    \centering
    \begin{tabular}{|c|c|c|c|}
        \hline 
        Valve & $T_{1}$ (s) at 10 G  & $\nabla B$ (G/cm) & $T_{1,wall}$ (s)\tabularnewline
        \hline 
        \hline 
        ILL & $970\pm90$ & no data & no data\tabularnewline
        \hline 
        1 & $760\pm40$ & no data & no data\tabularnewline
        \hline 
        2 & $960\pm30$ & 0$.0069\pm0.0004$ & $3170\pm540$\tabularnewline
        \hline 
        3 & $940\pm30$ & $0.0069\pm0.0005$ & $2490\pm400$\tabularnewline
        \hline 
        4 & $880\pm50$ & 0$.0079\pm.0007$ & $3450\pm1550$\tabularnewline
        \hline 
    \end{tabular}
    
\end{table}

All the tests with valves with Al KF style flanges, except valve 1, have $T_{1}$ values consistent with the glass stopcock. However, valve 1 can still be used on one of the dilution volumes where the gradient requirements are less stringent than at the polarization cell.

\subsection{Tests of MEOP valves with swage style connection}

The remaining three valves have swage style connections, and an adapter required to connect the swage connection to the glass flange on the test system. These adapters, made from Al 3003, are 20 in in length, with a KF style flange at one end. The aluminum tube is bent so it fits into the available space. The 20 in of tube is required for the actual system, although the flange is not. The flange was removed prior to the actual installation of the valve in the GHS. 3 valves of this style are required for the MEOP system. Table 2 shows the test results 

\begin{table}[H]
    \caption{Measured $T_{1}$ at 10 G for the 3 MEOP valves with swage style flanges. Additional measurements were taken between 5 and 15 G for all 3 valves,which were used to extract values for the gradient and wall $T_{1}$.The pressure in the cell for all tests was 1.0 Torr}
    \centering
    \begin{tabular}{|c|c|c|c|}
        \hline 
        Valve & $T_{1}$ (s) at 10 G  & $\nabla B$ (G/cm) & $T_{1,wall}$ (s)\tabularnewline
        \hline 
        \hline 
        5 & $1100\pm90$ & $0.0063\pm0.0003$ & $3230\pm500$\tabularnewline
        \hline 
        6 & $1080\pm70$ & $0.0073\pm0.0002$ & $5950\pm170$\tabularnewline
        \hline 
        7 & $1120\pm70$ & $0.0071\pm0.0003$ & $6370\pm650$\tabularnewline
        \hline 
    \end{tabular}

\end{table}

All the tests of valves with swage style flanges have $T_{1}$ values better than the test with the glass stopcock; clearly demonstrating the valves will not prevent achieving a 1000 s $T_{1}$ in the actual polarization system.

\section{Design and testing of MEOP coil}

\subsection{Description and design of the MEOP coil }

The MEOP system requires a highly uniform magnetic field in the \textsuperscript{3}He polarization direction. For a \textsuperscript{3}He cell at 1 Torr in a 5 G holding magnetic field, gradients need to be smaller than 0.0833\textbackslash\%/cm for a $T_{1}$ of 1000 s and less than 0.0263 \%/cm for 10,000 s. To accomplish this, an electromagnet was designed using the physical interpretation of the magnetic scalar potential as described in \cite{Crawford1}. For a region free of currents, the magnetic scalar potential $\phi$ satisfies $\nabla^{2}\phi=0$, which can be solved for a given magnetic field ($\vec{B}=\mu\vec{H}$), geometry and boundary conditions, resulting in magnetic scalar potential isosurfaces. The isocontours that are formed at the intersection of the isosurfaces and the geometrical boundaries represent the paths that electric current must follow in order to produce the magnetic intensity $\vec{H}=-\nabla\phi$. If the chosen set of isocontours differ by a fixed amount of magnetic scalar potential, $\Delta\phi$, then they share the same value of electric current $i\propto\Delta\phi$ and can be connected in series. An example of the application of this method can be found in \cite{Marissa}.

For the MEOP system it is necessary to provide a uniform magnetic field in a region of space large enough to accommodate the components of the optical system. In addition, the electromagnet has to confine the magnetic field and the corresponding return flux in a defined region of space so that so that interference with other components of the SOS system is avoided. COMSOL Multiphysics\textsuperscript{\textregistered} \cite{COMSOL} software was used to determine the magnetic scalar potential isocontours for the configuration shown in figure 
, resulting in a double cosine theta coil (main coil) with circular endcaps. 

\begin{figure}[H]
    \centering
    \includegraphics[width=1\textwidth]{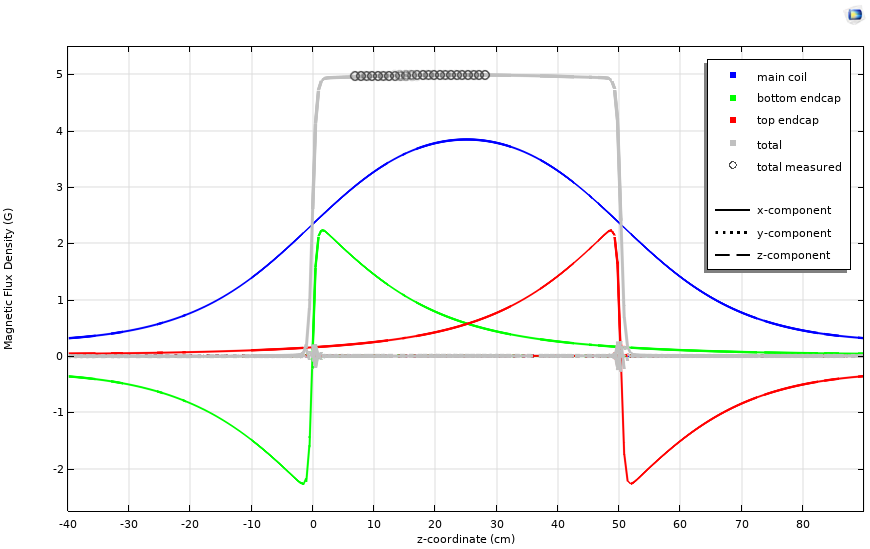}\caption{Magnetic flux density produced on axis by the main coil (blue), the bottom (green) and the top (red) endcaps, as well as the sum (gray). The $x$, $y$ and $z$ components are represented by solid, dotted and dashed lines, correspondingly. The $y$ and $z$ components are effectively zero.}
    \label{fig:on-axis-3-comp}
\end{figure}

\begin{figure}[H]
    \centering
    \includegraphics[width=1\textwidth]{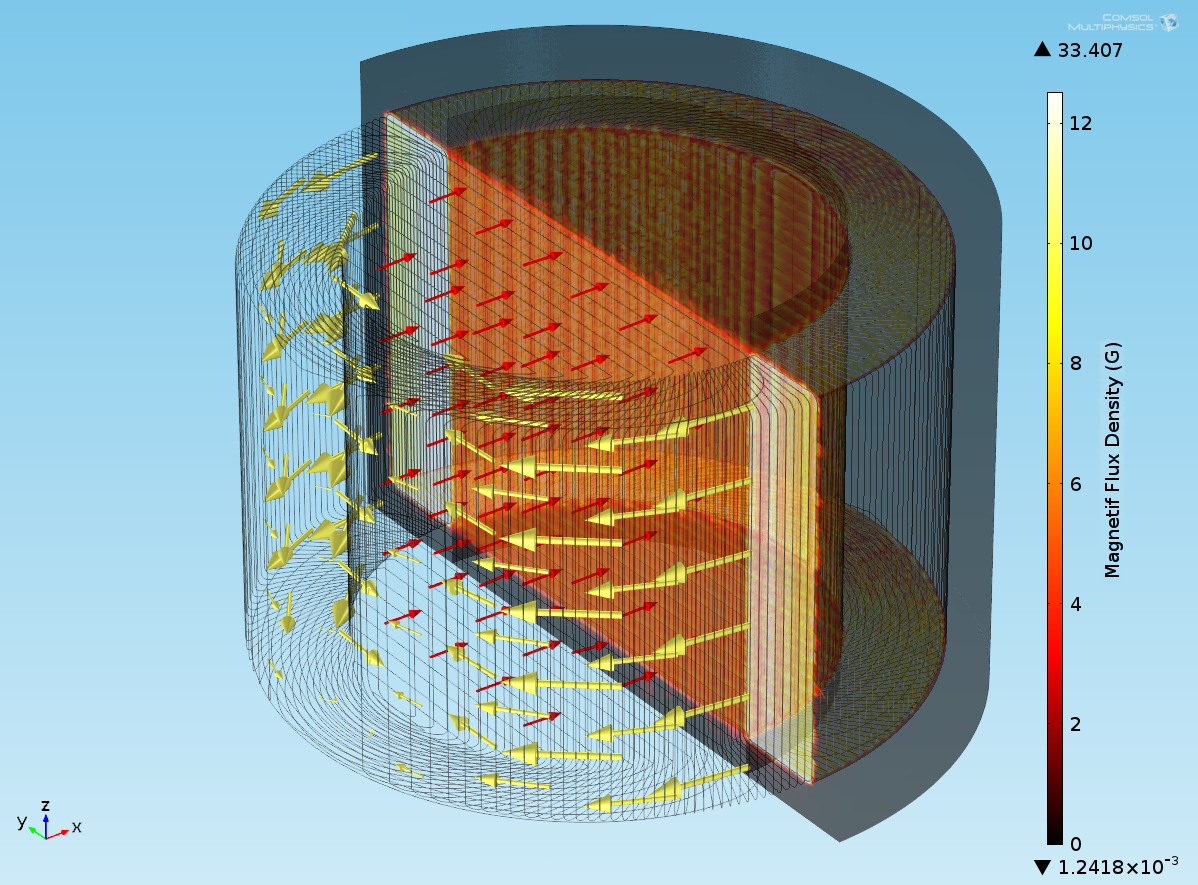}\caption{Magnetic flux density. Red arrows correspond to the flux in the inner region of the coil, while yellow ones correspond to the return flux between the inner and outter cylindrical walls of the main coil. The intensity is also shown with a color legend (G); a uniform 5 G magnetic field ($B_{0}$) is produced along the $x$-direction in the inner region, while there is zero magnetic field outside the outer cylindrical wall and endcaps.}
    \label{fig:flux-arrows}
\end{figure}

Figure \ref{fig:on-axis-3-comp} shows the three spatial components of the magnetic fields produced by the main coil and endcaps along the axis of the coil for both, simulated as well as measured field. As can be seen in figure \ref{fig:flux-arrows}, the electromagnetic device produces a uniform magnetic field in the $x$-direction in the inner region, while the return flux is contained between the inner and outer walls of the main coil; there is no magnetic field produced by the device outside the outer wall of the main coil or the endcaps.

The coil model was implemented by 3D printing ABS (Acrylonitrile Butadiene Styrene) structures with grooves for copper enameled magnet wire gauge 12, which properly withstands an electric current of 8 A at 16 V, corresponding to the chosen isocontours that produce a 5 G field intensity in the inner region of the coil. The dissipated power of the entire device (main coil and endcaps) is less than 128 W. A 7.6 cm wide by 43.2 cm tall window was placed on one side the main coil in order to allow for the laser beam into the optical system (see figure \ref{fig:MEOPManifoldCoil}). This window also allows for the pneumatic, vacuum, and RF connections to exit the coil.

\subsection{Measurements of the MEOP coil field and gradients}

\begin{figure}[H]
    \centering
    \includegraphics[scale=1]{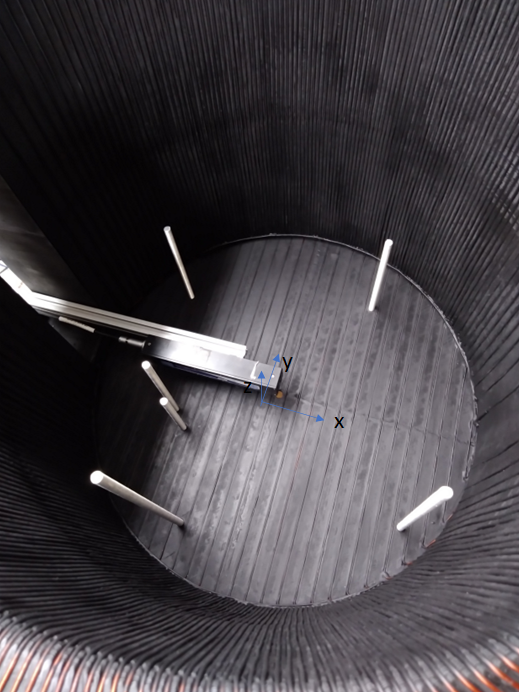}
    \caption{Setup for the field map of the interior of the coil with both endcaps attached and powered. The scan was conducted from 6 cm to 29 cm in the vertical ($z$) direction, 39 cm in the direction along the holding field ($x$) and 3 cm along the $y$ direction. The positions are measured relative to the center of the bottom endcap of the coil. The scan along the y direction is limited by the width of the hole and can only be performed once the probe had completely cleared the hole (-4 cm to 20 cm in $x$). The hole also limits the vertical scan to 6 cm or more above the bottom endcap}
    \label{fig:ScanSetup}
\end{figure}

Measurements of of the magnetic field produced by the coil were taken using a magnetic field mapper consisting of a 3 axis magnetometer, mounted on a 3 axis stage, located at the University of Kentucky. With the coil endcaps in place the only access into the coil is through the hole in the side of the coil for the laser and polarimeter. Scanning through this hole limits the region that can be scanned but allows scans to be taken with the coil complete and both endcaps powered. 
See figure \ref{fig:ScanSetup} for picture of the scan setup. 

The field was measured by first measuring the background field without the coil powered then repeating the measurement with the coil powered. By subtracting the background from the powered field the true field produced by the coil is determined. 

\begin{figure}[H]
    \centering
    \subfloat[]{\includegraphics[scale=0.47]{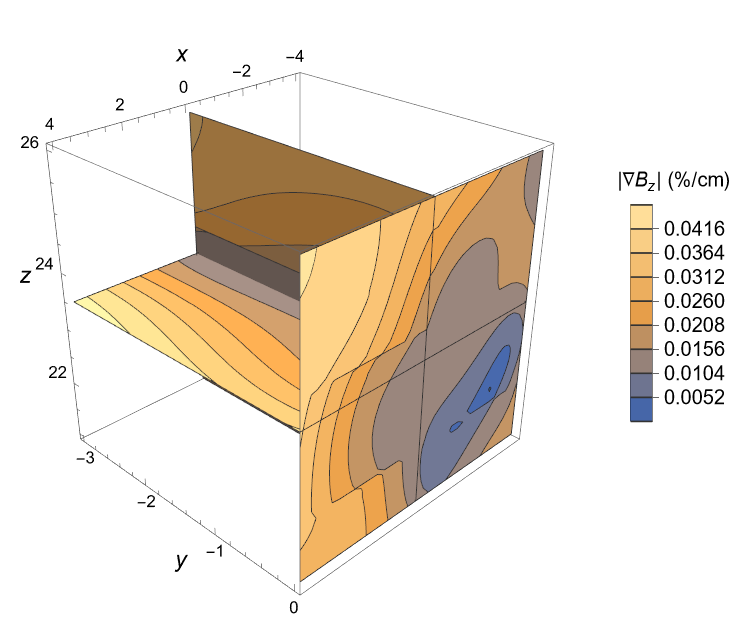}}\subfloat[]{\includegraphics[scale=0.47]{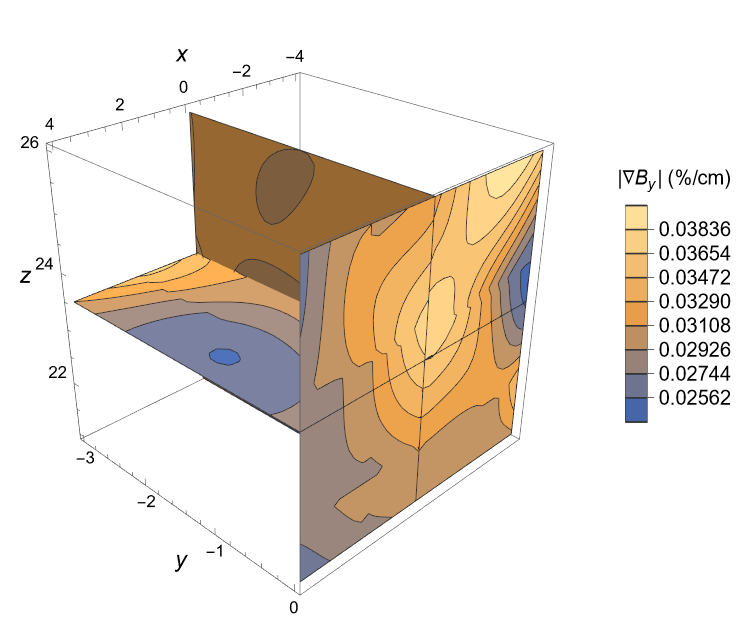}}

    \centering
    \subfloat[]{\includegraphics[scale=0.47]{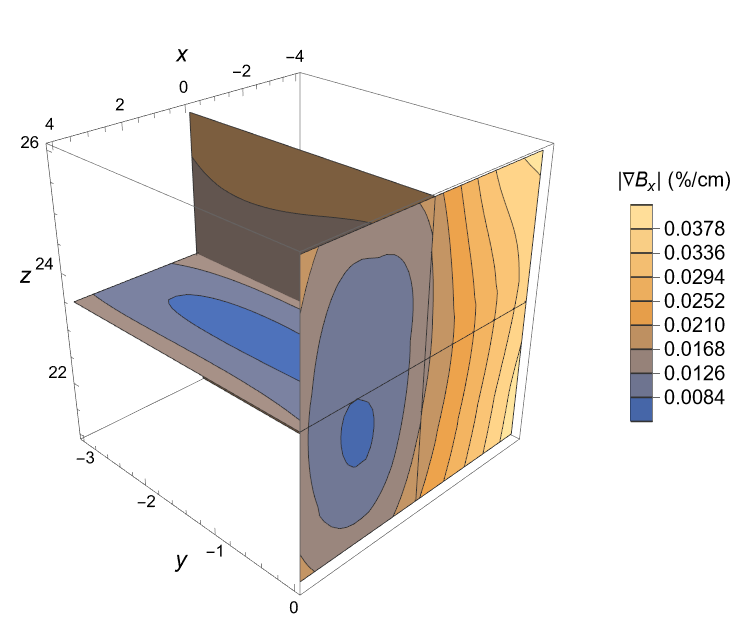}}

    \caption{Magnitude of the magnetic field gradients in the region around the MEOP cell. (a) shows $|\nabla B_{z}|$, (b) shows $|\nabla B_{y}|$, and (c) shows $|\nabla B_{x}|$. All gradients are less than 0.0833 \%/cm required for $T_{1}>1000$s.Due to the mounting of the probe only one side of the cell is scanned in the $y$ direction. Although the field should be symmetric in this direction. The plot shows contour plots through the the central planes of the cell assuming the cell is centered at (0,0,23.5 cm). The cylindrical cell is oriented along the $x$ axis, approximately 4.5 cm in diameter and 6 cm long.}
    \label{fig:CellGrad}
\end{figure}

The data can then be interpolated and the magnitude of the gradient calculated from the interpolation function. Given our cell is 6 cm long, 4.5 cm in diameter, is centered on the central axis of the coil, and is located approximated 23.5 cm above the bottom endcap it can be seen from figure \ref{fig:CellGrad} that the gradients at the cell location will be sufficient to achieve a $T_{1}>1000$s (transverse gradient<0.0833\%/cm). 

\begin{figure}[H]
    \centering
    \includegraphics[width=.95\textwidth]{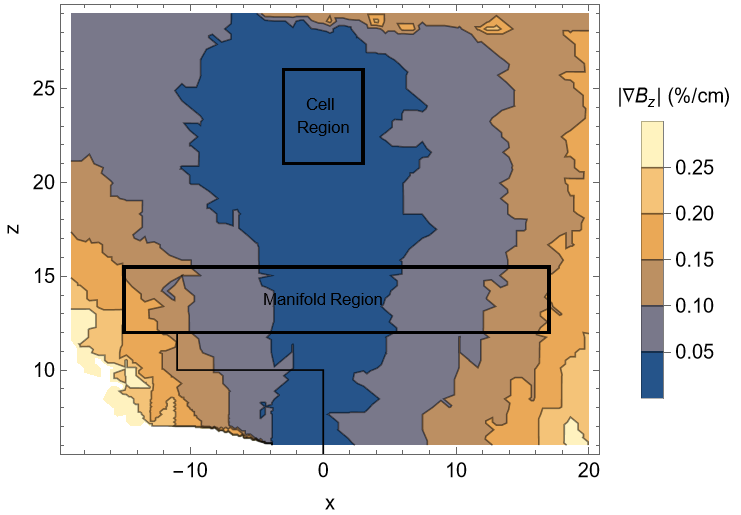}
    \caption{Magnitude of the $B_{z}$ gradient over a 2D $x$,$z$ slice of the cell at $y=0$. This slice covers the cell, manifold and the section of the injection line inside the coil as pictured. During injection the \protect\textsuperscript{3}He will briefly see gradients as high as 0.2 \%/cm or a $T_{1}$ of 180s (for P=1 Torr) . The $B_{y}$ gradient is below 0.0833 \%/cm over the entire region.}
    \label{fig:2DSlice}
\end{figure}

Additionally by examining the gradients along a 2D slice of the map extending across the MEOP manifold and injection line in the direction  of the holding field (figure \ref{fig:2DSlice}), we see that the $T_{1}$ times seen during injection are long compared to the injection time of a half a second. The timescale for injection is discussed in section 5. 

\begin{figure}[H]
    \centering
    \includegraphics{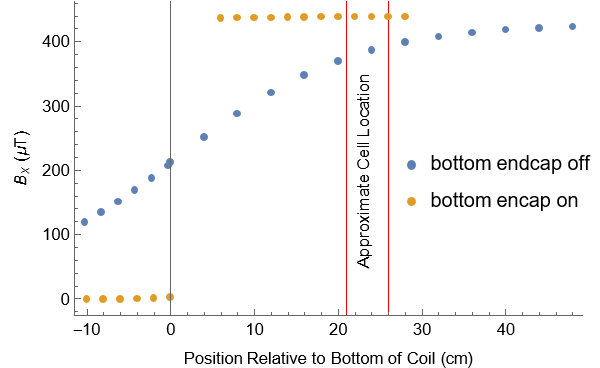}
    \caption{holding field produced by the MEOP coil along central axis of the coil. The bottom of the coil is located at 0 cm. If the bottom endcap is on the \textsuperscript{3}He see an abrupt 440 uT drop in the holding field over 6 cm, when exiting the coil. If the bottom endcap is off the transition when exiting the coil is much smoother but the field is no longer uniform in the cell region}
    \label{fig:CapPowerCompare}
    \end{figure}

The largest transverse gradient seen in this region is in  $B_{z}$ which is approximately \SI{0.2}{\percent/\centi\meter}, which corresponds to a $T_{1}$ of \SI{180}{\second} (assuming a pressure of 1 Torr) . However, this analysis only considers the gradients seen while inside the coil. Upon exiting the coil through the bottom cap the holding field produced by the coil drops to nearly zero (See figure \ref{fig:CapPowerCompare}). In order to preserve the polarization, the holding field needs to be extended along the injection path to join with the holding field produced by the SOS $B_{0}$ coil (the holding field for the NMR measurements in superfluid field measurement cell), so the \textsuperscript{3}He have a non zero holding field through the entire injection process. 

It is possible to accomplish this by partially depowering, or turning off the bottom endcap completely, prior to injection. In this case the falloff in the holding field is greatly smoothed. however, the gradients at the cell increases to 1.3\% in the case of a completely depowered bottom endcap. Additionally, the gradient just outside the bottom of the coil (0 to -2~cm) is 4.2\% increasing to 5.2\% 10 cm away from the bottom. These gradients where calculated assuming 22 uT background field along the N/S oriented holding field (Earth's field N/S component at 36N -79E). Figure \ref{fig:CapPowerCompare} shows $\vert \nabla B_{z} \rvert$ along the central axis of the coil calculated using an interpolation of the measured field. Section 5.2 contains of the effect these gradients would have of the polarization during injection. Figure \ref{fig:CapPowerCompare} compares the change in the holding field along the central axis of the coil, both with and without the bottom endcap powered. 

\begin{figure}[H]
    \centering
    \includegraphics{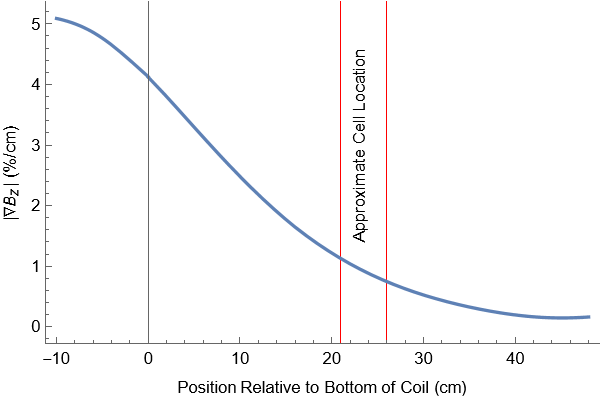}
    \caption{$\vert \nabla B_{z} \rvert$ along central axis of MEOP coil. The bottom of the coil is located at 0 cm. Gradients are calculated as a percentage of the total field (field produced by coil plus estimated background) along the x axis (holding field direction). The background field is estimated to be 22 uT in this direction, based on the magnitude and direction of Earth's field.}
    \label{fig:Gradalongz}
    \end{figure}

\subsection{Comparison of simulation and measurement}

It is also useful to compare the measured field and gradients of the coil as built to the simulated field and gradients of the coil design. The uniformity of the magnetic flux density in the inner region of the coil is presented in table \ref{tab:Uniformity} in the form of residuals $\lvert\langle B-\langle B\rangle\rangle / \langle B\rangle\rvert$ along 5 and 10 cm regions from the center of the coil. Residuals for both, the simulated and the measured fields are presented.

\begin{table}[H]
    \caption{Calculated and measured magnetic flux density residuals evaluated in a 10 and 5 cm radius region from the center of the coil.}
    \centering
    \renewcommand{\arraystretch}{1.5}
    \begin{tabular}{|r|c|c|}
    \cline{2-3}
    \multicolumn{1}{c|}{ } & \multicolumn{2}{c|}{Residuals (\%)}\\
    \cline{2-3}
    \multicolumn{1}{c|}{ } & Simulated & Measured\\
    \hline
    @ 10 cm & 0.0598 & 0.1281\\
    \hline
    @ 5 cm & 0.0152 & 0.0315\\
    \hline
    \end{tabular}

    \label{tab:Uniformity}
\end{table}

Similarly, gradients are presented in table \ref{tab:Gradients}. The gradients presented are the average gradient between the center of the coil and a radial region 5 to 10 cm away where it can be seen that the 0.0833\%/cm limit is satisfied for a 10 cm radius region centered at the center of the coil, while the 0.0263\%/cm limit is basically satisfied in a 5 cm radius region.

\begin{table}[H]
    \caption{Gradients. All values are below the 0.0833\%/cm limit for a 1000 s T1; values above the 0.0263\%/cm limit for 10 000 s T1 are shown in red.}
    \centering
    \renewcommand{\arraystretch}{1.5}
    \begin{tabular}{|r|c|c|c|c|}
    \cline{2-5}
    \multicolumn{1}{c|}{} & \multicolumn{4}{c|}{Gradients (\%/cm)} \\
    \cline{2-5}
    \multicolumn{1}{c|}{} & \multicolumn{2}{c|}{@ 5 cm} & \multicolumn{2}{c|}{@ 10 cm}\\
    \cline{2-5}
    \multicolumn{1}{c|}{} & Calculated & Measured & Calculated & Measured \\
    \hline
    \multicolumn{1}{|c|}{$\frac{\partial B_x}{\partial x}$} & 1.9168$\times 10^{-4}$ & -0.00901 & 1.5007$\times 10^{-4}$ & -0.01995 \\
    \multicolumn{1}{|c|}{$\frac{\partial B_x}{\partial y}$} & 7.8343$\times 10^{-4}$ & 0.01067 & 0.00151 & -0.00135 \\
    \multicolumn{1}{|c|}{$\frac{\partial B_x}{\partial z}$} & -0.00466 & -0.00342 & -0.00831 & 0.01490 \\
    \multicolumn{1}{|c|}{$\frac{\partial B_y}{\partial x}$} & -0.00129 & 0.00820 & -0.00229 & 0.01034 \\
    \multicolumn{1}{|c|}{$\frac{\partial B_y}{\partial y}$} & -0.00167 & \color{red}0.02913 & -0.00161 & \color{red}0.02755 \\
    \multicolumn{1}{|c|}{$\frac{\partial B_y}{\partial z}$} & 1.7454$\times 10^{-4}$ & -0.00222 & -6.7855$\times 10^{-4}$ & -0.00367 \\
    \multicolumn{1}{|c|}{$\frac{\partial B_z}{\partial x}$} & 0.00662 & 0.01071 & 0.01056 & 0.01191 \\
    \multicolumn{1}{|c|}{$\frac{\partial B_z}{\partial y}$} & 0.00308 & 0.00148 & 0.00373 & 0.00927 \\
    \multicolumn{1}{|c|}{$\frac{\partial B_z}{\partial z}$} & 0.00148 & 0.01727 & 0.00146 & 0.00175 \\
    \hline
    \end{tabular}

    \label{tab:Gradients}
\end{table}

\section{\texorpdfstring{\textsuperscript{3}}{3}He injection from room temperature into the LHe}

\subsection{\texorpdfstring{\textsuperscript{3}}{3}He injection simulations using COMSOL\texorpdfstring{\textsuperscript{\textregistered}}{R}}
In order to determine the heat load into the measurement cell during injection as well as the polarization loss, a COMSOL\textsuperscript{\textregistered} model of 2 mm diameter Kapton injection line, with appropriate heat sinking, was constructed and used to model the flow of helium from the room temperature 80 cc polarization cell into the superfluid helium. The 80 cc cell itself was not modeled but rather the mass flow rate of the approximately 5 mg helium gas out of the cell and into the Kapton capillary was used as a boundary condition for the top of the tube. 

\subsubsection{Estimating the mass flow rate from the polarization cell}

Assuming an ideal gas the rate which atoms in the polarization cell hit section of the cell's surface ($dS$) is given by

\begin{equation}
    \frac{dN_{\mathrm{cell}}}{dt} = -\frac{dS}{4V}\sqrt{\frac{8 k_b T}{\pi m}}N_\mathrm{cell}
\end{equation}

Where $V$ is the cell's volume, $m$ is the mass of \textsuperscript{4}He, and $N_{cell}$ is the number of atoms currently in the cell. We approximate the gas as being completely composed of \textsuperscript{4}He as the \textsuperscript{3}He is only 1.3 parts in 500. If we treat the cell as being directly connected to the Kapton capillary then the rate of atoms entering the capillary, with cross section $S$, can be found by integrating of the cross sectional area giving

\begin{equation}
    \frac{dN_\mathrm{cell}}{dt} = -\frac{S}{2V}\sqrt{\frac{8 k_b T}{\pi m}}N_\mathrm{cell}
\end{equation}

The solution is a decaying exponential of the form

\begin{equation}
    N_\mathrm{cell} = N_0\,e^{{-t/\tau}}
    \label{11}
\end{equation}

where $N_{0}$ is the initial number of atoms in the cell, and $\tau$ is a characteristic timescale. Rewriting in terms of mass flow by multiplying both sides by $m$ we have

\begin{equation}
    M_\mathrm{cell} = M_0\,e^{{-t/\tau}}
    \label{12}
\end{equation}

With $M_0$ equal to the initial mass of helium (approximately 5 mg) initially in the cell, and the characteristic time for the atoms to leave the polarization cell $\tau$ given by

\begin{equation}
    \tau = \frac{4V}{S}\sqrt{\frac{\pi m}{8 k_bT}}
    \label{13}
\end{equation}

 It should be noted that this model ignores any gas that may return back into the volume, the impedance of the opening between the two volumes, or any other effects. It is essentially an estimate for the gas escaping into an infinite vacuum through area S, unimpeded by any frictional or thermal factors.
 
Substituting our geometric parameters into the equation above, we find that the characteristic timescale for the mass flow is \SI{0.081}{s}. We use this value in the COMSOL\textsuperscript{\textregistered} simulation that follows.

\subsubsection{Simulation details}

Based on the calculation in the previous section we estimate the mass flow through the top of the Kapton tube as

\begin{equation}
    \frac{dM_\mathrm{tube}}{dt}=\frac{1}{\tau}M_{0}e^{-t/\tau}\frac{11}{10}(1-e^{-10t/\tau})
    \label{14}
\end{equation}

\begin{figure}[H]
    \centering
    \includegraphics[width=30em]{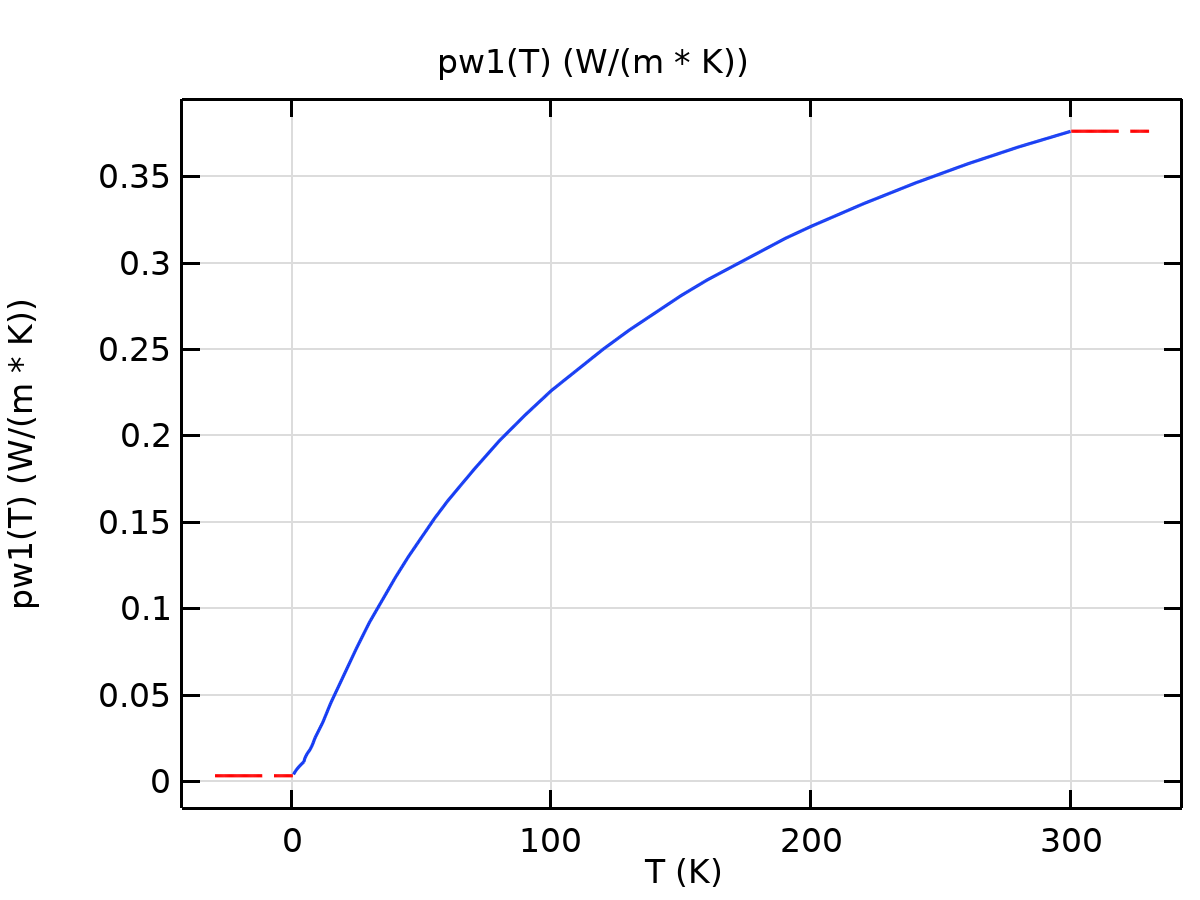}
    \caption{Kapton conductivity used for COMSOL\textsuperscript{\textregistered} simulations. Conductivity data above 5 K is from \cite{Rule:1990ht}, while data below 5 K is from \cite{Lawrence}.}
    \label{fig:conductivity}
\end{figure}

This is derived from eq. (\ref{12}) with an extra factor of $\frac{11}{10}\left(1-e^{-10t/\tau}\right)$ included to ensure smooth turn on of the mass flow. The 11/10 is needed to properly normalize mass flow. Additionally, the top of the tube is modeled with a constant temperature boundary condition (293.15 K). The outlet is modelled as a pressure boundary condition, representing the vapor from the superfluid film at the low temperature end of the tube. This pressure is taken to be 5 mbar, although the vapor pressure at lambda point of superfluid helium is approximately 50 mbar \cite{donnelly}. This is done to avoid numerical instabilities observed in the simulation at larger pressure boundary conditions. Additionally, setting the boundary condition also sets the initial pressure distribution along to the tube to the boundary value. It is unlikely the tube will be entirely filled with 50 mbar of gas. Furthermore, preliminary simulations done at different pressure boundary conditions show similar final results to the 5 mbar case. The tube itself is modelled as a thin layer boundary condition for heat flow with thickness 0.1 mm. The thermal conductivity data for Kapton HN are collected from \cite{Rule:1990ht} for the regime above 5 K, and from \cite{Lawrence} for the regime below 5 K. A plot of the thermal conductivity data is shown in figure \ref{fig:conductivity}. Specific heat data are taken from \cite{NISTcryo}. The 4 K helium bath is modelled as a temperature boundary condition on the external side of the thin layer boundary condition from $z=0.3$ m to $z=1.3$ m. To reduce the likelihood of numerical instabilities, the initial temperature distribution is chosen to be 4K from $z=0$ m to $z=1.3$ m, with a linear ramp in temperature from 4 K to 293.15 K from $z=1.4$ m to $z=1.5$ m. The fluid dynamics is simulated using the algebraic y+ turbulence model in COMSOL\textsuperscript{\textregistered}.

\subsubsection{Simulation results}

The latent and sensible heat flux are estimated from the temperature and flow rate of the helium gas at the outlet boundary condition. The latent heat is determined by multiplying the mass flux across the outlet by the heat of vaporization of helium, 90 J/mol; thus the total latent heat is independent of pressure or flow rate provided that the total mass flux remains the same across all simulations and condensation along the pipe walls is neglected. The sensible heat is defined as the heat needed to cool the gas at the outlet to 2 K at constant pressure. 
\begin{figure}[H]
    \centering
    \includegraphics[width=30em]{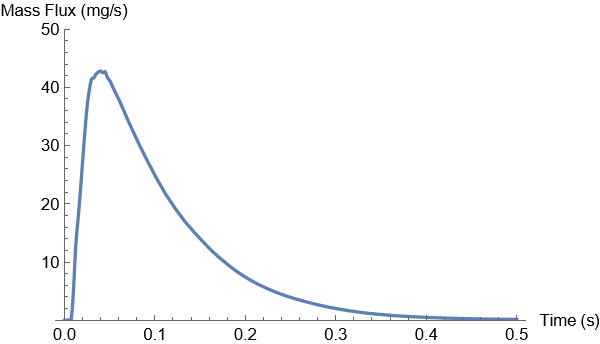}
    \caption{Mass flux of helium gas across Kapton tube outlet.}
    \label{fig:flux}
\end{figure}
The mass flux is shown in figure \ref{fig:flux}, while the cumulative heat is shown in figs. \ref{fig:5mbar}. The sensible heat is small (0.02 J) compared to the latent heat, while the latent heat is approximately 0.11 J. If a boundary pressure of 40 mbar is used instead, the latent heat is still dominant and equal to 0.11 J, although the sensible heat is approximately 0.05 J.
\begin{figure}[H]
    \centering
    \includegraphics[width=30em]{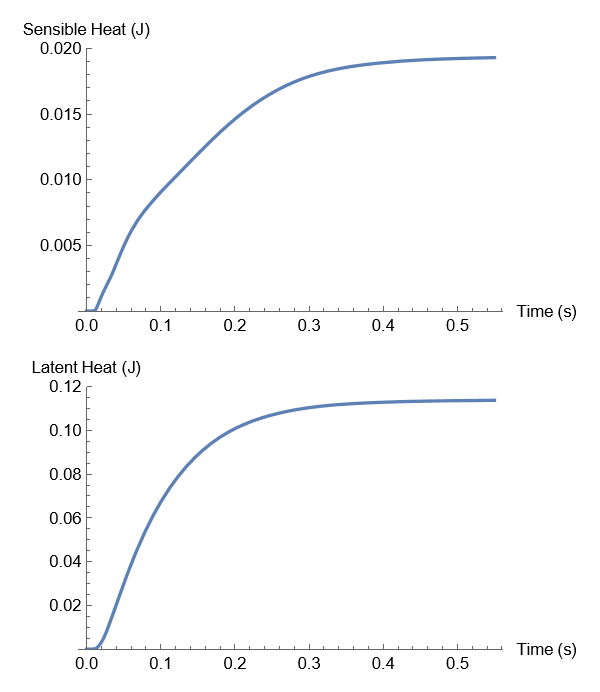}
    \caption{Cumulative heat versus time for 5 mbar outlet pressure.}
    \label{fig:5mbar}
\end{figure}

\subsection{Estimate of polarization loss during injection}

The amount of polarization loss during injection must be understood in order to obtain the desired polarized \textsuperscript{3}He concentration in the measurement cell. Gradients along the injection path, and at the cell, after turning off the bottom endcap of the MEOP holding field coil will result in depolarization during injection. Results from the \textsuperscript{3}He injection COMSOL\textsuperscript{\textregistered} simulations are used in estimating the polarization loss 

\subsubsection{Polarization loss at cell}

When calculating the polarization loss at the cell during injection we assume the bottom endcap has been turned off and therefore the transverse gradient at the cell is 1.3 \%/cm (See section 4.2). Additionally, we use eq. (\ref{12}) for the mass of helium in the cell as a function of time. The cell is at room temperature, and if we assume the validity of the ideal gas law in the cell during injection, we can now calculate the pressure in the cell as a function of time, and use eq. (\ref{1}) to estimate $T_{1}$ in the cell, giving the rate of depolarization at the cell as

\begin{equation}
    \frac{dN}{dt}=x_{0,3}\frac{M_\mathrm{cell}(t)}{m_\mathrm{He}}\frac{e^{-t/T_{1}(P(t))}}{T_{1}(P(t))}
    \label{15}
\end{equation}

where $P(t)$ is the pressure as a function of time given by $P(t)=\frac{M_\mathrm{cell}(t)k_{b}T}{m_\mathrm{He}V_\mathrm{cell}}$, $m_\mathrm{He}$ is the mass of a helium atom, and $V_\mathrm{cell}$ is the volume of the MEOP cell (80 cc), and $x_{0,3}$ is the initial concentration of polarized \textsuperscript{3}He in the cell.

\begin{figure}[H]
    \centering
    \includegraphics[width=30em]{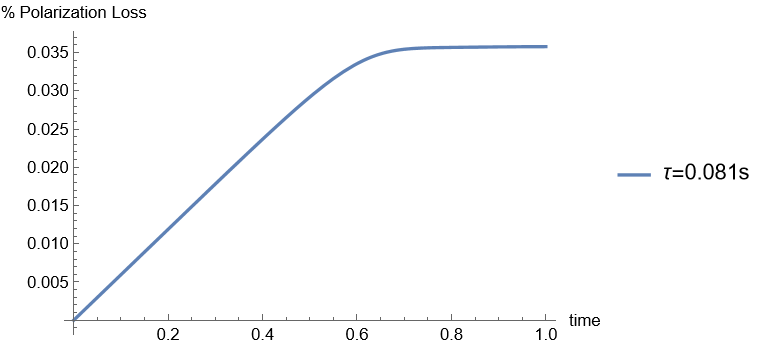}
    \caption{Predicted polarization loss at the Polarization cell during injection for $\tau=0.081$ s, calculated in section 5.1.1, for a transverse gradient of 1.3\%/cm}
    \label{PolCellLoss}
\end{figure}

Numerically integrating this as a function of time we find that over the injection period about \SI{0.037}{\percent} of the initial polarization is lost. Figure \ref{PolCellLoss} shows the polarization loss at the cell during injection.

\subsubsection{Polarization loss along injection path}

To estimate the polarization loss due to gradients along the injection path during injection we consider the probability of a particle depolarization during the time it takes to travel from the cell to the liquid surface. The gradient along the path is assumed to be a constant value, and the $T_{1}$ is calculated using the average pressure along the tube as a function of time as determined from the 5 mbar outlet pressure injection simulation. Given the way this simulation is constructed we cannot distinguish between the atoms initially in the tube vs the helium injected from the cell. As such this calculation only considers the polarization loss due to gradients, and doesn't consider the possibility of polarized helium remaining in the tube after injection. Thus the main effect of using a larger value for the pressure boundary condition is a longer $T_{1}$ due to the higher average pressure. We also take the time between the beginning of injection and the maximum pressure at the liquid level for the 5 mbar outlet pressure simulation as a typical travel time for the injected \textsuperscript{3}He. For $\tau=0.081$ this corresponds to 0.04 s. Doing so we estimate the cumulative fraction of polarized helium reaching the bottom of the tube as

\begin{equation}
    \centering
    \frac{M_{pol}(t)}{M_{0}}=\intop_{0}^{t}\frac{dt'}{M_{0}}\frac{dM_\mathrm{cell}}{dt'}e^{-T_\mathrm{travel}/T_{1}(P_\mathrm{ave}(t'))}
    \label{16}
\end{equation}

where $T_{travel}$ is the time between the beginning of injection and the maximum pressure at the tube liquid level, and $P_{ave}(t)$ is the average pressure along the tube as a function of time. It should also be noted we are evaluating the average pressure at the time the helium enters the tube. If we were to instead evaluate the the pressure at $P(t'+T_{travel})$, the time the particle reaches the bottom instead, the polarization loss in the tube increases by 0.09 \%. The cumulative fraction of \textsuperscript{3}He depolarized in the tube during injection as a function of time can then be expressed as the difference between the fraction of helium that has entered the tube and the fraction of polarized helium that reaches the liquid level giving

\begin{equation}
    \centering
    \frac{M_\mathrm{depol}(t)}{M_{0}}=\frac{1}{M_{0}}\intop_{0}^{t}\frac{dM_\mathrm{cell}}{dt'}dt'-\frac{M_\mathrm{pol}(t)}{M_{0}}
    \label{17}
\end{equation}

\begin{figure}[H]
    \centering
    \includegraphics[width=30em]{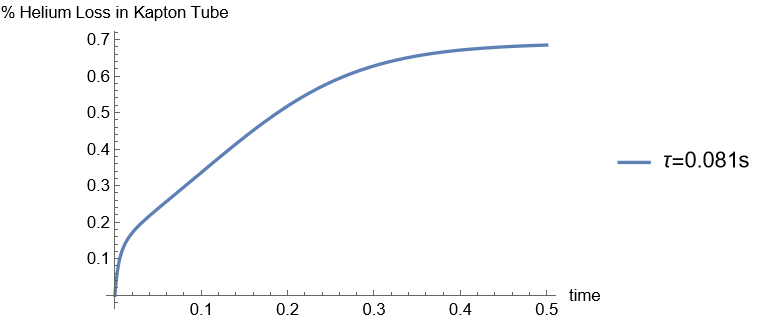}
    \caption{Predicted polarization loss in the tube due to gradients during injection for $\tau$ between 0.064 s and $\vert \nabla B_{z} \rvert=5.2$ \%/cm}
    \label{fig:PolTubeLoss}
\end{figure}

Additionally, the measured field gradient 10 cm below the MEOP coil (Figure \ref{fig:Gradalongz}) is used for the gradient over the entire injection path. Figure \ref{fig:PolTubeLoss} shows the predicted cumulative polarization loss due to gradients in the tube as a function of time during injection. Resulting in a predicted polarization loss of 0.68 \% in the tube. Combining this result with the polarization loss in the cell we than have a predicted total polarization loss due to gradients of 0.72 \%.

\section{Conclusions}

In the present publication we describe a MEOP setup capable of production of 80\% polarized $^3$He, diluting it in a wide rage of relative concentrations with $^4$He, and injecting the mixture into the cryogenic cell cooled down to below 500 mK temperature.  This setup is a part of the Systematic and Operational Studies apparatus, which is a cryogenic NMR test-bed developed in the frame of the cryogenic nEDM project ~\cite{Ahmed} based on the Golub and Lamoreaux proposal~\cite{golub94}. The wide range of relative $^3$He concentrations and operational temperatures in 100--500~mK are required to study the pseudomagnetic field, the trajectory correlation functions of the \textsuperscript{3}He and neutrons that will determine the size of the nEDM systematic errors, and to develop precise spin manipulations techniques including spin dressing~\cite{cianciolo2024sos}. Possible applications of the SOS apparatus to axion searches similar to the search described in \cite{Guigue} is also being explored. The concentration and temperature choice is based on the condition that $^3$He interaction with phonons prevail the $^3$He - $^3$He collisions.

The polarization and injection system required for the SOS is unique in that it is a room temperature MEOP based polarization and injection system capable of delivering polarized \textsuperscript{3}He into a superfluid filled measurement cell at concentrations as low as $10^{-10}$ and with minimal losses. Previous work with an room temperature MEOP based injection system in \cite{Yoder} was limited to an estimated 10 \% polarization and concentrations of $10^{-4}$ or greater. Similarly systems exhibiting higher final polarizations using circulation methods \cite{Hayden1} have similarly, to our knowledge, only been demonstrated at concentrations too large to study the \textsuperscript{3}He-phonon interactions of interest for the nEDM experiment. 

Our design utilizes a glass manifold that is used to dilute the polarized \textsuperscript{3}He before injection in order to deliver the desired final concentration in the measurement cell. As well a system to pressurize the \textsuperscript{3}He with \textsuperscript{4}He between each round of dilution and before injection. This is critical to preserving the the polarization of the gas given the pressure dependence of the $T_{1}$. The manifold is enclosed by a 5 G holding field for MEOP, provided by a double cosine theta coil. The field generated by the coil has a field uniformity better than 0.0833 \%/cm at the polarization cell. Since minimizing the field gradients is essential to provide a $T_{1}\approx1000$ s the 8 pneumatically actuated valves inside the field region were tested to ensure they didn't introduce any additional magnetic gradients. These tests of the \textsuperscript{3}He "friendliness" of the valves indicated that any gradients introduced by the valves were at the same level as borosilicate glass and therefore negligible (the $T_{1}$ at the polarization cell in a test system, where a pneumatic valve is used to isolate the cell is the same as an all glass system). The valve under test is closed during relaxation measurements. 

Additionally,simulations indicate that it will take less than 1 s to inject the the gas into the cell with the design described, and injection will introduce a heat load of approximately 0.13 J. This heat load is dominated by condensing the helium into the liquid. Furthermore, polarization loss during injection due to gradients is estimated to be 0.72 \% in the case of unrestricted flow in the 2 mm diameter, 1.5 m long injection line with a 5.2 \%/cm field gradient. Such minimal loss of polarization during injection will allow us to be able to deliver almost all of polarization produced at the polarization cell into the measurement cell, maximizing the NMR signal and minimizing neutron absorption.

\acknowledgments{This work was supported in part by the US Department of Energy under grant number DE-FG02-97ER41042 and through subcontract number 4000196635 administered through Oak Ridge National Laboratory.}

\bibliographystyle{ieeetr}
\bibliography{references}

\end{document}